# Zero-dead Time NMR in Low Field: Single-Sideband Technique Revisited

Baosong Wu[1,2]

1 Department of Radiology and Biomedical Imaging, Yale University, New Haven, CT 06520, USA

2 ABQMR, Albuquerque, NM 87106, USA

**Abstract:** Recently the rapid-scan technique is reviving in NMR or EPR, because of its benefits of zero-dead time and low RF power. While signal baseline is still a big problem in such experiments, time-share method has been used to indirectly avoid it. However, it is obviously not a truly zero-dead time method. Other data-processing methods were also adopted to deal with raw data. Here we try to use single-sideband technique at 11.4MHz to mitigate this obstacle. The prospect is that single sideband technique can be used in rapid-scan experiment for low/high field imaging.

**Introduction:** Pulsed NMR/MRI has been widely used since the advent of mini-computers that made Fourier transform NMR possible [1], because of its advantage in high sensitivity compared to the classic continuous-wave(*cw*) NMR [2]. In normal *cw* experiments, magnetization is flipped a relatively small angle and observed in steady-state, which leads to its low sensitivity. The signal with wiggles was first reported by Bloembergen et al. in their experiments [3]. A technique called rapid-scan NMR was proposed in 1974[4], which showed higher sensitivity than *cw* experiments. Recently people begin to revive the rapid-scan technique in NMR [5,6] or EPR [7,8], because of its benefits in zero-dead time and low RF power. Such technique makes imaging with zero-dead time possible in low field. The advantage of super-low RF power makes it possible develop clinical MRI in high field (such as 7T or even higher) and avoid specific absorption ratio(SAR) limitation. While signal baseline is still a big problem in rapid-scan experiments, time-share method has been used to indirectly avoid it [5]. However, it is obviously not a truly zero-dead time method. Other data-processing methods were also adopted to deal with raw data [7]. Here we try to use single-sideband technique [9] to mitigate this obstacle.

**Theory and Methods:** The single sideband technique, using both sweeping and modulation fields, provides almost complete freedom from base line drift. When field is swept during experiment, quadrature hybrid or magic tee needs to be used while one terminal is connected to probe and a balanced probe or 50 Ohm resistant connect to another terminal. Any change in probe will cause base-line drift but modulation field can mitigate this drift.

The resonance signal can be obtained when the static magnetic field, $B_0$, is swept through Larmor frequency (as shown in Fig.1(a)). Here the modulation field(Fig.1(b)) is mixed with the magnetic field so that the instantaneous magnetic field at the sample is $B_0+B_m \cos(\omega_m t)$, thus, $B_m$ is the peak amplitude of the field modulation and $\omega_m$ is the angular frequency of the field modulation. The detected signal will contain audio component as well as DC components. Any change of balance in the probe will change the DC level and leave the AC component unaffected. Thus, by selecting only the AC component of the signal, the effect of base-line drift can be minimized.

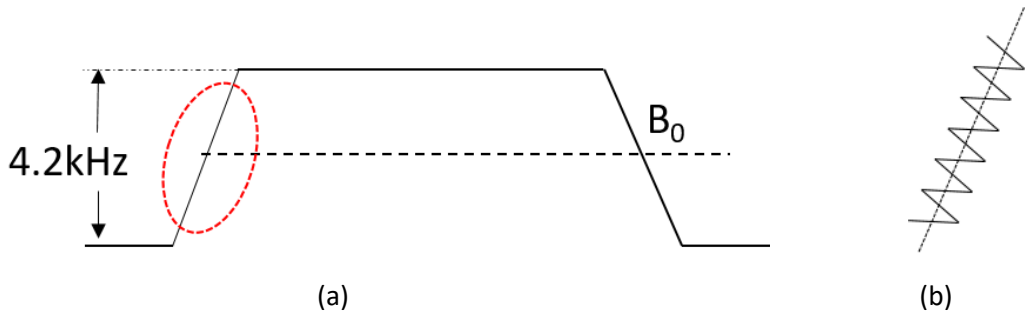

(a)  (b)

Fig.1 Field-sweep and field-modulation waveforms. (a) depicts the trapezoidal waveform for a linear sweeping field. The B₀ dotted line shows the resonance frequency at the center of sweep field. (b) The solid line shows the modulation waveform after mixing the sweep field in the red-circle region (a).

Figure 2 shows the system schematic for single sideband modulation. We used a 100mm gap 0.267 Tesla permanent magnet and a continuous RF source (tens of watts). The RF power was separated into two parts by a splitter, one channel passed the 10dB attenuator to magic tee, and the other was used as a reference source to mix with the output of pre-amplifier. Two mixers were used, one for removing RF frequency and the other for modulation frequency.

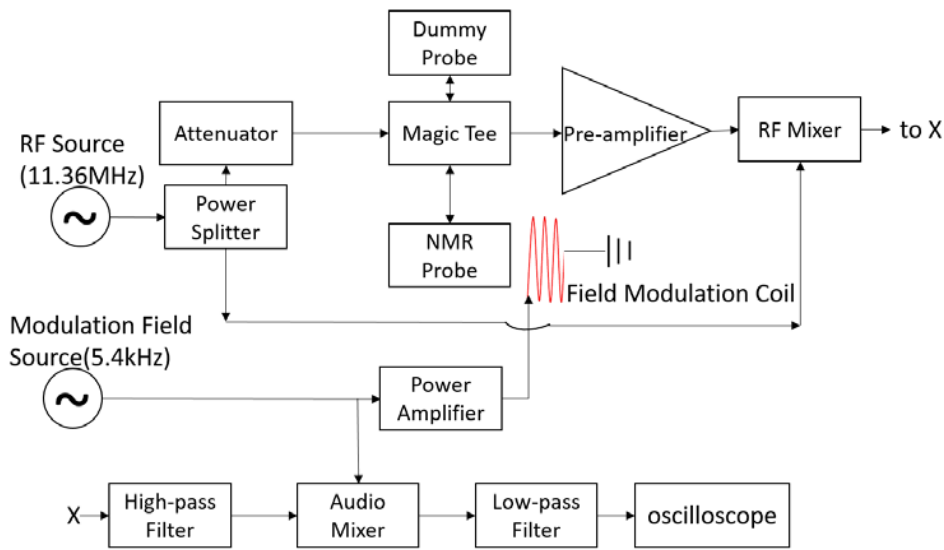

Fig.2 Single sideband system including modulation field and sweeping field.

The solenoid RF coil consisted of 12 turns of AWG30 copper wire with a radius of 8mm in a shielded aluminum box. The modulation coil, oriented with its plane parallel to the axis of RF coil, consisted of 25 turns of AWG38 copper wire with a radius of 30mm. It generated modulation field, which was parallel to B₀ field. Peanut oil (T2*=5ms) in glass tube was positioned in the RF coil, and the resonance region was close to the center of the modulation coil. We used a low-frequency power amplifier to drive the modulation coil. The field-sweep coil, not shown in Fig.2, was mounted on one pole of the magnet and driven by a programmable power supply. A low-noise preamplifier was used to obtain better signal gain after the output of low-pass filter. The signal was then displayed on the oscilloscope.

**Results:** The amplitude $B_m$ of modulation field was measured by a pickup coil, and the ratio of $\gamma B_m/\omega_m$ was ~2. The field-sweep range is 4.2kHz in ~50ms, so $df/dt$ = 82kHz/s. Figure 3 shows two channels with channel 1 being the *cw* signal which corresponds to the field-sweep function in channel 2.

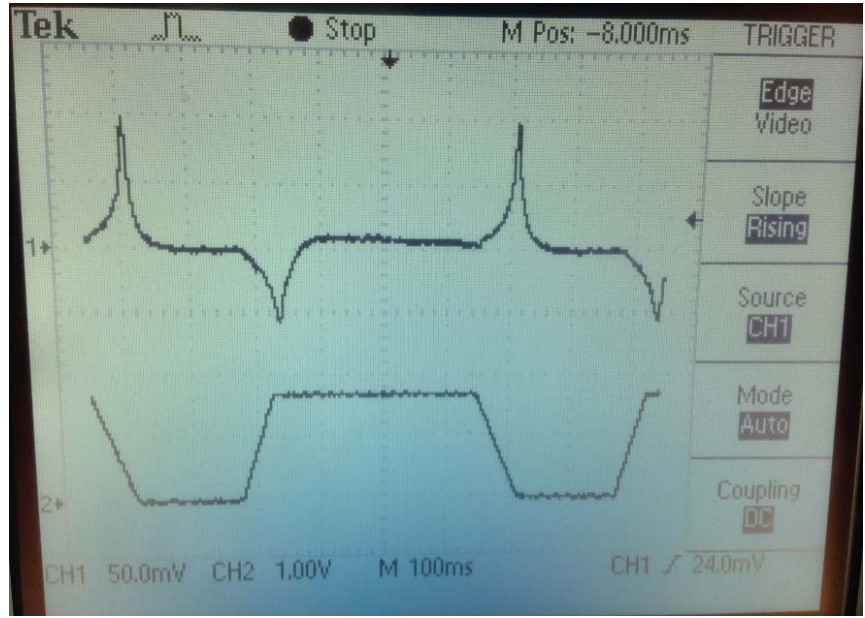

Fig.3 Signal (CH 1) displayed, together with the sweep wave form (CH 2).

**Conclusion:** *cw* signal was obtained which verifies that this technique can effectively solve the base-line drift. A more powerful driver for field-sweep coil is needed to realize much larger sweeping range and satisfy the rapid-scan condition ($\left|\frac{B_1}{dB_0/dt}\right| \ll \sqrt{T_1 T_2}$). We confirm that single sideband technique can be used in rapid-scan experiments to achieve zero-dead time and low SAR imaging.

Acknowledge: The author appreciates discussion with Dr. Eiichi Fukushima and Mark S. Conradi, and also thanks Prof. R. Todd Constable for his support.